\documentclass{Interspeech}

\interspeechcameraready

\title{MOPSA: Mixture of Prompt-Experts Based Speaker Adaptation \\for Elderly
Speech Recognition}

\author[affiliation={1}]{Chengxi}{Deng}
\author[affiliation={2*}]{Xurong}{Xie}
\author[affiliation={1}]{Shujie}{Hu}
\author[affiliation={3}]{Mengzhe}{Geng}
\author[affiliation={2}]{Yicong}{Jiang}
\author[affiliation={1}]{Jiankun}{Zhao}
\author[affiliation={1}]{Jiajun}{Deng}
\author[affiliation={1}]{Guinan}{Li}
\author[affiliation={1}]{Youjun}{Chen}
\author[affiliation={1}]{Huimeng}{Wang}
\author[affiliation={1}]{Haoning}{Xu}
\author[affiliation={1}]{Mingyu}{Cui}
\author[affiliation={1*}]{Xunying}{Liu}

\affiliation{}{The Chinese University of Hong Kong}{Hong Kong SAR, China}
\affiliation{Institute of Software}{Chinese Academy of Sciences}{China}
\affiliation{}{National Research Council Canada}{Canada}
\email{cxdeng@se.cuhk.edu.hk, xurong@iscas.ac.cn, xyliu@se.cuhk.edu.hk}


\usepackage{comment}
\usepackage{multirow}
\usepackage{pifont}
\newcommand{\cmark}{\ding{51}}
\newcommand{\xmark}{\ding{55}}
\usepackage{graphicx}
\usepackage[table]{xcolor}
\usepackage{bm}
\usepackage{amsmath,amsfonts,amssymb,mathtools}
\usepackage{cite}
\usepackage[utf8]{inputenc}
\usepackage[titletoc,title]{appendix}
\usepackage{makecell}
\usepackage{tipx}

\usepackage{color}
\usepackage{multirow}
\usepackage{hhline}

\begin{document}
\maketitle
\renewcommand{\thefootnote}{*}%
\footnotetext{Corresponding author.}%
\renewcommand{\thefootnote}{\arabic{footnote}}%

\begin{abstract}
This paper proposes a novel Mixture of Prompt-Experts based Speaker Adaptation approach (MOPSA) for elderly speech recognition. It allows zero-shot, real-time adaptation to unseen speakers, and leverages domain knowledge tailored to elderly speakers. Top-K most distinctive speaker prompt clusters derived using K-means serve as experts. A router network is trained to dynamically combine clustered prompt-experts. Acoustic and language level variability among elderly speakers 
are modelled using separate encoder and decoder prompts for  Whisper. Experiments on the English DementiaBank Pitt and Cantonese JCCOCC MoCA elderly speech datasets suggest that online MOPSA adaptation outperforms the speaker-independent (SI) model by statistically significant word error rate (WER) or character error rate (CER) reductions of \textbf{0.86\%} and \textbf{1.47\%} absolute (\textbf{4.21\%} and \textbf{5.40\%} relative). Real-time factor (RTF) speed-up ratios of up to \textbf{16.12} times are obtained over offline batch-mode adaptation. 
\end{abstract}

\section{Introduction}

In the current aging society, ensuring effective communication for the elderly becomes increasingly vital for maintaining their social engagement and quality of life. Elderly speech is often characterized by both imprecise articulations stemming from weakened neuromotor control and linguistic degradation associated with cognitive decline \cite{becker1994natural_dbank}. The resulting lower speech intelligibility not only impedes the daily communication of the elderly population but also poses wider societal implications, such as reduced social 
participation. 
Since current foundation models for automatic speech recognition (ASR) primarily target normal speakers\cite{whisper,baevski2020wav2vec,hsu2021hubert,chen2022wavlm}, their application to elderly speech remains a challenging task\cite{hu2022exploring,ssl_shujie_taslp}. Therefore, it is essential to investigate how to effectively adapt these foundation models to better accommodate elderly speech recognition.

\par
Elderly speech presents multifaceted challenges to existing deep learning-based ASR technologies, including:
{\textbf{1) data sparsity}}\cite{geng2024homogeneous,ssl_shujie_taslp} due to the difficulty in collecting large-scale datasets from elderly speakers with mobility limitations;
and \textbf{2) speaker heterogeneity}\cite{geng2024homogeneous,geng2022speaker} among elderly speakers, where typical sources of variability in speech, such as accent and gender, are further compounded by varying degrees of phonetic deterioration and linguistic expression degradation.
The advent of large-scale foundation models\cite{hsu2021hubert,chen2022wavlm,whisper,baevski2020wav2vec} has further intensified these challenges as their massive parameters require extensive data to prevent overfitting, yet research on their application in the elderly speech domain remains limited \cite{ssl_shujie_taslp,hu2024structured}.
\par
To address the aforementioned challenges in elderly speech recognition, various speaker adaptation methods have been proposed\cite{geng2022speaker,cuhk_elderly_zi_ye,geng2024homogeneous}. Earlier studies focused on specially designed speaker-dependent (SD) parameters to model speaker variability. More recently, adapter-based speaker parameter estimation approaches have been investigated within large-scale foundation models for elderly speech recognition~\cite{hu2024structured}, while research in this direction remains limited.

\begin{figure*}[h]
    \centering    \includegraphics[width=0.8\textwidth]{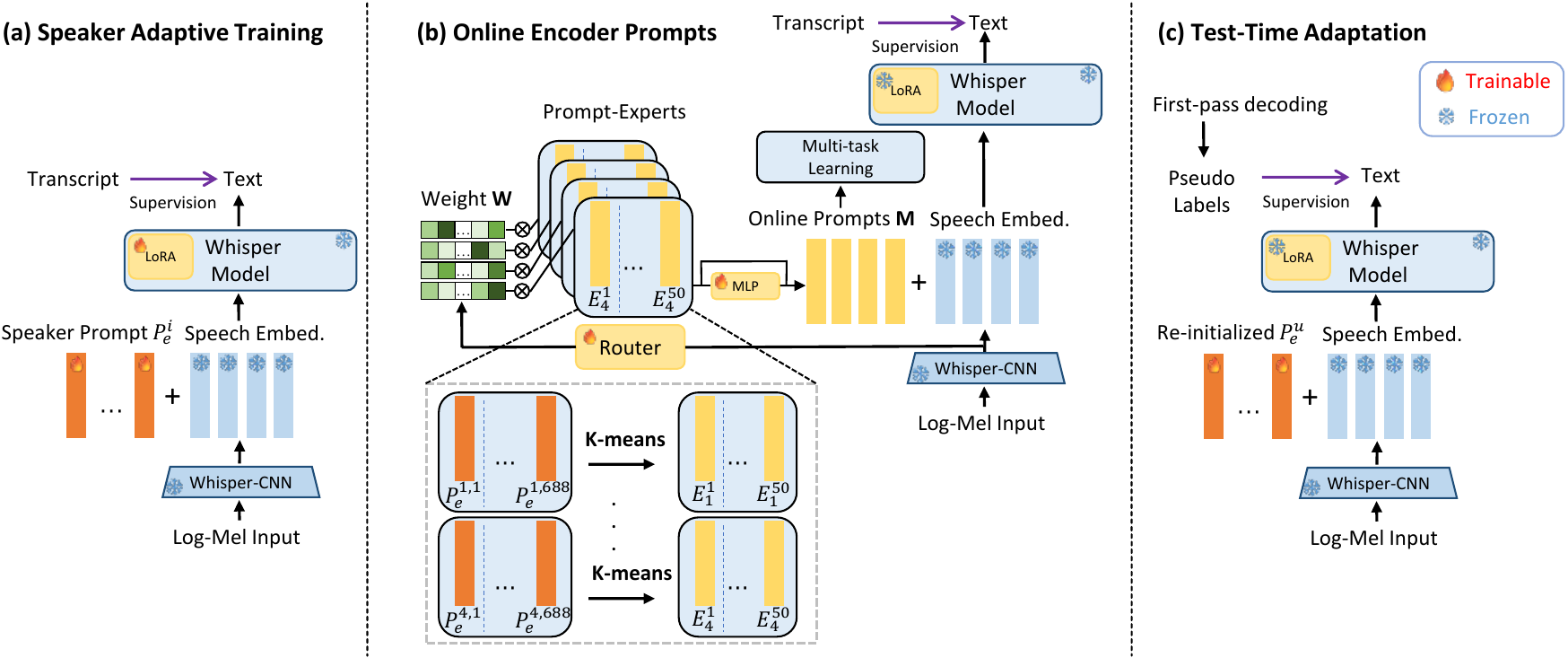}

    \caption{Examples of \textbf{(a)} the speaker adaptive training, 
    \textbf{(b)} the online mixture of prompt-experts adaptation,
    and \textbf{(c)} the batch-mode speaker prompts test-time adaptation performed on the encoder of Whisper.} 

\label{fig:pipeline_method}
\vspace{-0.5cm}
\end{figure*}

In contrast, more advanced speaker adaptation techniques targeting normal speech have been explored beyond conventional approaches\cite{gemello2007linear,li2010comparison,tan2015cluster,zhang2016dnn,zhao2016low,samarakoon2016subspace,li2018speaker,xie2019fast,fan2019speaker,huang2021rapid,sim2021robust,deng2023confidence,jiajun_high_performance}. These methods can be divided into two broad categories:
{\textbf{1) Mixture-of-Experts (MoE)}}\cite{jacobs1991adaptive_moe,jordan1994hierarchical_moe}, where researchers have leveraged multiple sets of Low-Rank Adaptation (LoRA)\cite{lora} or adapters\cite{adapter_tuning}, each acting as an expert specialized in capturing distinct information. The experts and their weights are the SD parameters and they can be parameterized to represent speaker-specific features or domain-specific information pertinent to the speaker\cite{hu2024structured,zhao24_saml_interspeech}.
{\textbf{2) Prompt-based adaptation}}\cite{liu2021gpt}, which trains a small set of speaker-dependent prompts or transformation to preserve the model’s knowledge and reduce computational overhead\cite{jiang24b_interspeech}.
\par
However, the current MoE-LoRA-based and prompt-based adaptation approaches suffer from several limitations when directly applied to elderly speech recognition:
\textbf{1) Limited capacity for generalization: }
The SD parameters estimated from training speakers often fail to generalize to unseen speakers due to differences between the training and testing speakers. Limited data from elderly individuals further biases the estimation of elderly speaker parameters, weakening zero-shot adaptation for unseen speakers.
\textbf{2) Latency:} Generating pseudo-labels and performing test-time adaptation introduces additional delays hindering effective real-time communication for the elderly.
\textbf{3) Lack of adaptation to language deficiency:}
Existing approaches primarily focus on acoustic adaptation, overlooking the language-specific characteristics of elderly speech, such as repetitive phrasing and disfluencies.\par
To address these limitations, this paper proposes a novel Mixture of Prompt-Experts based Speaker Adaptation approach (MOPSA) for elderly speech recognition,
which enables zero-shot adaptation for unseen speakers, operates in real-time, and leverages domain knowledge tailored to elderly speech.
The proposed MOPSA consists of three key components, including: 
\textbf{1)} A set of speaker prompts is first derived for each speaker via speaker adaptive training\cite{anastasakos1996compact_sat}, and the K-means-based cluster method is further applied to cluster these speaker prompts into a set of prompt-experts, this clustering-based approach enables better generalization to unseen speakers; \textbf{2)} A router network is trained to dynamically combine these prompt-experts, enabling efficient online adaptation; 
\textbf{3)} In addition to encoder prompts, we also use decoder-level speaker prompts to capture non-acoustic, language-level diversity among elderly speakers. 
\par
The main contributions of this work are as follows: \\
\textbf{1)} To the best of our knowledge, this paper presents the first study on the use of Mixture of Prompt-Experts based speaker adaptation for elderly speech recognition. This can be summarized in three key points: \textbf{a)} While previous research overlooked adaptation capabilities for unseen speakers, our prompt-experts demonstrate enhanced robustness and better generalization to unseen speakers. \textbf{b)} Compared to batch-mode adaptation methods, our routing mechanism dynamically combines prompt-experts in real-time, achieving high performance while maintaining low latency. \textbf{c)} Compared to previous speaker adaptation methods that only focus on acoustic variability~\cite{geng2022speaker,geng2024homogeneous,hu2024structured}, we explore decoder-level speaker prompts to further capture language-level diversity among elderly speakers.\\ 
\textbf{2)} Experimental results on the DementiaBank Pitt \cite{becker1994natural_dbank} and JCCOCC MoCA \cite{jccocc_datasets} elderly speech datasets suggest that the proposed method outperforms the SI model by statistically significant word error rate (WER) or character error rate (CER) reductions of \textbf{0.86\%} and \textbf{1.47\%} absolute (\textbf{4.21\%} and \textbf{5.40\%} relative). Real-time factor (RTF) speed-up ratios of up to \textbf{16.12} times are obtained over offline batch-mode adaptation.

\begin{table*}[h]
\caption{Performance contrast of the proposed online MOPSA and batch-mode prompt-based speaker adaptation with different comparable methods on DementiaBank Pitt and JCCOCC MoCA. ``Inv." and ``Par." refer to clinical investigators and elderly participants. Enc, Dec, and Enc\&Dec denote the application of speaker prompts to the encoder, decoder, and both encoder-decoder, respectively. $^\ast$ denote statistically significant (MAPSSWE \cite{gillick1989some}, $\alpha$ = 0.05) improvements obtained against the SI baseline ASR systems (Sys.1)}
    \centering
    \vspace{-0.35cm}
    \resizebox{\linewidth}{!}
    {
    \begin{tabular}{c|c|c|c|c|c|cc|cc|c|c|c|c|c|c|c}
    \hline\hline 
    \multirow{3}{*}{Sys.} & 
    \multirow{3}{*}{Model} & 
    \multirow{3}{*}{\shortstack {Speaker\\Modeling}} & 
    \multirow{3}{*}{\shortstack {SAT}} & 
    \multirow{3}{*}{\shortstack {TTA}} & 
    \multirow{3}{*}{Online} & 
    \multicolumn{7}{c|}{DementiaBank Pitt WER(\%)} &
    \multicolumn{4}{c}{JCCOCC MoCA CER(\%)} \\
    \cline{7-17} 
    & & & & & & 
    \multicolumn{2}{c|}{Dev.}&\multicolumn{2}{c|}{Eval.} & \multirow{2}{*}{All}&\multirow{2}{*}{\shortstack {SD Parm.}}&\multirow{2}{*}{RTF}&
    \multirow{2}{*}{Dev.} & \multirow{2}{*}{Eval.}
    & \multirow{2}{*}{All}&\multirow{2}{*}{{SD Parm.}}\\
    \cline{7-10}
    & & & & & &Par.&Inv.&Par.&Inv.&&&&&&&\\
    \hline\hline

1&\multirow{12}{*}{\shortstack {Whisper\\(LoRA)}} & - & - & - & - & 28.79 &	12.76 & 	20.68 &	12.65 &	20.43 & -&0.24&28.68&25.79 &27.23 & - \\
\cline{3-17}

2 & & LHUC & \cmark&\multirow{5}{*}{\cmark}&\multirow{5}{*}{-}& 28.45&12.40&20.11&11.43&20.02$^\ast$&0.15M&4.15&30.77&27.63&29.19&0.1M\\
3 & & RAB\cite{hu2024structured} & \xmark & & & 29.54 &13.14&21.27&12.87&20.99&0.53M&2.78&29.35&26.55&27.94&0.53M\\

{\cellcolor{cyan!15}4} & &Prompt Enc& \cmark & & & 27.53$^\ast$&12.44&19.53$^\ast$&12.43&\textbf{19.60$^\ast$}&0.60M&4.03&27.50$^\ast$&24.32$^\ast$&\textbf{25.90$^\ast$}&0.4M\\

{\cellcolor{orange!20}5} & & Prompt Dec&\cmark &&&28.12&12.16$^\ast$&20.01&12.76&\textbf{19.81$^\ast$}&0.15M&3.25&27.75$^\ast$&24.49$^\ast$&\textbf{26.09$^\ast$}&0.1M\\

6 & & Prompt Enc\&Dec&\cmark &&
&27.41$^\ast$&12.05$^\ast$&19.25$^\ast$&11.76&\textbf{19.33$^\ast$}&0.75M&4.13&27.23$^\ast$&24.15$^\ast$&\textbf{25.69$^\ast$}&0.50M\\

\cline{3-17}
       
7 & & i-vector & \cmark & \multirow{6}{*}{\xmark} & \multirow{6}{*}{\cmark} &29.32 & 13.13 &21.75 &	10.99 & 20.92&\multirow{6}{*}{-}&0.27&39.04&36.16&37.59&\multirow{6}{*}{-}\\

8 & & x-vector & \cmark &  &  & 31.49 & 14.96 & 23.37 & 12.87&22.84&&0.27&29.87&27.43&28.64& \\

9 & & ECAPA-TDNN\cite{wang2024advancing}& \cmark & & &29.01 & 13.85 & 21.27 & 10.54 & 20.88&&0.27&33.48&30.19&31.83\\

{\cellcolor{yellow!50}10} & & MOPSA Enc & \cmark &  &  &28.49 & 12.32 & 19.48$^\ast$ & 10.88& \textbf{19.88$^\ast$}&&0.25&28.10$^\ast$&25.10$^\ast$&\textbf{26.59$^\ast$}&\\

11 & & MOPSA Dec & \cmark &  &  &28.76 &12.85 & 20.22 &11.10& 20.33&&0.25&27.54$^\ast$&25.20$^\ast$&\textbf{26.36$^\ast$}&\\
12 & & MOPSA Enc\&Dec & \cmark &  &  &27.64$^\ast$& 12.52 & 19.08$^\ast$ &11.21& \textbf{19.57$^\ast$}&&0.27&27.15$^\ast$&24.39$^\ast$&\textbf{25.76$^\ast$}   &\\

\hline\hline
\end{tabular}
}
\label{tab:table_online}
\vspace{-0.2cm}
\end{table*}

\vspace{-5pt}
\section{Large-Scale Foundation Model Whisper}
\vspace{-5pt}

Whisper\cite{whisper} is a transformer-based model capable of processing multiple languages and handling various speech-related tasks. The input to Whisper is log-Mel spectrogram $\bm{X} \in \mathbb{R}^{D \times T}$, where $D$ and $T$ respectively denote the feature dimension and length of the input. The convolutional block downsamples the input by a factor of 1/2, 
which is subsequently transformed by the encoder into hidden representations $\bm{H}_{e}$.
The decoder then auto-regressively generates text tokens $\bm{\hat{y}}$
conditioned on the encoder feature, previous tokens, and special tokens. Let $\hat{y}_{m}$ denote the current token and $\hat{{y}}_{1:m-1}$  the preceding token sequence. The aforementioned process can be formulated as:
\begin{equation}
\begin{aligned}
\bm{H}_{e} &= \text{Encoder}(\bm{\text{Conv}({\bm{X}})})\\
{\hat{y}}_{m} &= \text{Decoder}(\bm{s}, {\hat{y}}_{1:m-1}, \bm{H}_{e})\\
\end{aligned}
\end{equation}
where $\bm{s}$ denotes the special token sequence. The most common special token sequence is as follows: \texttt{"<|PREV|>}, \textit{decoder prompt}, \texttt{<|SOT|>}, \texttt{<|LANGUAGE|>}, \texttt{<|TRANSCRIBE|>}, \texttt{<|NO-TIMESTAMP|>"}. The \textit{decoder prompt} enables users to obtain personalized outputs by defining their own prompts.

\vspace{-3pt}
\section{Batch-Mode Speaker Prompt Adaptation}
\par
\noindent
\textbf{Adaptation Data Accumulation:} In batch-mode adaptation, speaker prompts are estimated using first-pass decoding ${\hat{\bm{Y}}}$  as pseudo-labels generated by the SI system, which requires substantial data from the testing speaker. This process introduces latency and is highly dependent on the quantity and quality of the pseudo-labels.

\par
\noindent
\textbf{Test-Time Adaptation :}
As shown in Fig. \ref{fig:pipeline_method}(c), during test-time adaptation (TTA), the parameters of the entire Whisper model remain frozen. 
For each testing speaker, a set of trainable prompts is initialized on both the encoder and decoder, which are then optimized using pseudo-labels as supervision.
Considering $U$ testing speakers, the process for incorporating the encoder-side prompts can be formulated as follows:
\vspace{-4pt}
\begin{equation}
\begin{aligned}
\bm{H}_{e}^{u} &= \text{Encoder}(\text{Concat}[\bm{P}^u_{e}, \bm{\text{Conv}(\bm{X})}])    
\end{aligned}
\end{equation} 
where $u\in\bm{U}$ indexes the testing speaker, $L_e$ denotes the length of the encoder-side prompts, $\bm{P}^u_{e} \in \mathbb{R}^{D \times L_{e}}$ indicates the encoder-side speaker prompts, and $\bm{H}_{e}^u\in \mathbb{R}^{D\times (L_{e}+T/2)}$ represents the concatenated hidden states. For incorporating the decoder-side speaker prompts, the process is formulated as:
\vspace{-3pt}
\begin{equation}
\begin{aligned}
    \hat{{y}}_{m} &= \text{Decoder}(\bm{s}(\bm{P}^u_{d}),\hat{{y}}_{1:m-1}, \bm{H}_{e}^u)\\
\end{aligned}
\end{equation} 
where $\bm{P}_{d}^u \in \mathbb{R}^{D \times L_{d}}$ denotes the learnable decoder prompts with length $L_d$ in the special token sequence $\bm{s}$. 
During test-time adaptation, pseudo-labels $\bm{\hat{Y}}$ are used as supervision to update both the encoder-side and decoder-side prompts. 
Let $\bm{P}^u=\{\bm{P}^u_{e},\bm{P}^u_{d}\}$  
denote the prompts for both sides. The optimization process is formulated as:
\vspace{-1pt}
\begin{equation}
\begin{aligned}
\bm{\hat{{P}}}^u = \underset{\{\bm{P}^{u}\}} {\arg\min}\{\mathcal{L}_C(\bm{\hat{Y}}^{u}|\bm{X}^u;\bm{P}^{u})\}
\end{aligned}
\end{equation}
where $\bm{X}^u$ denotes the log-Mel spectrogram for speaker $u$ and $\mathcal{L}_C$ represents the cross-entropy loss.
\par
\noindent
\textbf{Speaker Adaptive Training:}
As shown in Fig. \ref{fig:pipeline_method}(a), a set of trainable speaker prompts is initialized for each training speaker, following the standard speaker adaptive training (SAT) \cite{anastasakos1996compact_sat} process. For $I$ training speakers, the speaker prompts are represented as $\bm{P}^i=\{\bm{P}^i_{e},\bm{P}^i_{d}\}$ where $i\in\{1,2,...,I\}$ indexes the training speaker. Let $Y$ denote the ground truth transcripts and $\bm{r}$ represent the LoRA parameters shared among all speakers. 
The optimization process can be formulated as: 
\vspace{-5pt}
\begin{equation}
\{\bm{\hat{P}}^i, \hat{\bm{r}}\} = \underset{\{\bm{P}^{i}, \bm{r}\}}{\arg\min} \{\mathcal{L}_C (\bm{Y}^{i}|\bm{X}^i;\bm{P}^{i}, \bm{r})\}
\end{equation} 
\vspace{-2pt}
where ${\bm{X}^i}$ denotes the input log-Mel spectrogram of speaker $i$ and ${\bm{Y}^{i}}$ denotes the corresponding transcript.
\vspace{-3pt}
\section{Online Mixture of Prompt-Experts Speaker Adaptation}
\vspace{-1pt}
\textbf{MoE using speaker prompts:}
To enable online speaker adaptation for unseen speakers, as shown in Fig. \ref{fig:pipeline_method}(b), K-means clustering is applied to the speaker prompts from SAT, producing robust and generalizable prompt-experts. For $I$ training speakers, the encoder-side prompts for each speaker are represented as $\bm{P}_{e}^{i,l}\in \mathbb{R}^{D \times L_{e}}$, 
where $l\in\{1,2,...,{L}_{e}\}$ denotes the prompt position index. To generate prompt-experts, ${L}_{e}$ distinct K-means modules are utilized, while each K-means module clusters speaker prompts into $\bm{C}$ clusters.
Let $\bm{P}_{e}^l=\{\bm{P}_{e}^{1,l}, \bm{P}_{e}^{2,l},...\bm{P}_{e}^{I,l}\}$ denote the collection of prompts at the $l^{th}$ position. The corresponding expert for the $c^{th}$ cluster, $\bm{E}_{c}^{l}\in\mathbb{R}^{D}$, is given as:
\vspace{-8pt}
\begin{equation}
\begin{aligned}
    {\bm{E}}_{c}^{l} &= {\text{K-means}}(\bm{P}_{e}^l)\\
\end{aligned}
\end{equation}
The construction of prompt-experts on the decoder side follows the same approach as above.
\par
\noindent
\textbf{Router network:} To achieve real-time adaptation and enable generalization to unseen speakers, we propose a router network that dynamically combines prompt-experts to generate consistent speaker prompts. The router network on the encoder side consists of two main components: a global context network and a downsampling network. 
The \textbf{global context network} leverages standard multi-head attention layers\cite{attention} to model contextual relationships. The \textbf{downsampling network} consists of three sequential blocks\footnote{Each sequential block consists of three components: 1) CNN-1d layer with kernel size 5 and padding 2. 2) BatchNorm1d and 3) AvgPool1d layer with kernel size 2.} followed by linear layers. 
After passing through these sequential blocks, the features are reshaped and processed by linear layers, followed by a softmax operation to obtain the output $\bm{W}$ of
router network, where $\bm{W}=[\bm{W}^1,\bm{W}^2,...,\bm{W}^L]\in\mathbb{R}^{C\times L}$ and $\bm{W}^l=[w^l_1,w^l_2,...,w^l_C]^T\in\mathbb{R}^{C}$. The output $\bm{W}$ represents the weights for prompt-experts, and are then used to compute the weighted speaker prompts $\bm{Z}$ by combining the corresponding experts. Let $\bm{Z} = [\bm{Z}^1, \bm{Z}^2, \dots, \bm{Z}^L] \in \mathbb{R}^{D \times L}$, $\bm{Z}$ is transformed by a linear layer $\phi$ to obtain the online prompts $\bm{M}\in\mathbb{R}^{D \times L}$, and the process is expressed as follows:
\vspace{-9pt}
\begin{equation}
\begin{aligned}
{\bm{Z}^l = \sum_{j=1}^{\bm{C}} {w}_j^l \cdot \bm{E}_j^l}, ~~~~~
{\bm{M} ={\bm{Z}}+\phi(\bm{Z})}
\end{aligned}
\end{equation}
The online router network on the decoder side
shares the same architecture as described above, 
with the exception that its input is the acoustic features processed by the Whisper encoder.

\par
\noindent
\textbf{Multi-task Learning:}
To capture both local utterance-level details and  global speaker information, the training loss of the router network consists of three parts:
\textbf{1)}  The online prompts are integrated into the network and optimized using the cross-entropy loss $\mathcal{L}_{ASR}$ the same as Whisper's original training process\cite{whisper}, which is further adopted as the default loss for all online systems.
\textbf{2)} To ensure consistency in speaker identity, average pooling is applied to the online prompts $\bm{M}$, while the pooled representations are fed into a speaker recognition module optimized with a cross-entropy (CE) loss $\mathcal{L}_{Spkr}$. 
\textbf{3)} To guarantee consistency between the online prompts $\bm{M}$ generated by prompt-experts and those from SAT, a mean squared error (MSE) loss $\mathcal{L}_{MSE}$ is applied to align the online prompts with the SAT prompts. The overall router network cost function is given as  
$\mathcal{L}_{Router} = \mathcal{L}_{ASR}+\alpha\mathcal{L}_{Spkr}+\beta\mathcal{L}_{MSE}\footnote{$\alpha$ and $\beta$ for the encoder-side and decoder-side prompts are empirically set to 0.1, 0.02 for Table \ref{tab:table_online} Sys. 10, 11 and 0.1, 0.2 for Table \ref{tab:table_online} Sys. 12.}$. 
\vspace{-1pt}
\section{Experiments}
\par
\noindent
\textbf{Task description:}
The English \textbf{DementiaBank Pitt} \cite{becker1994natural_dbank} corpus is the most widely used publicly available elderly speech corpus for speech-based Alzheimer’s Disease (AD) diagnosis. It contains 33 hours of audio recordings from 292 interviews between elderly participants and clinical investigators during AD assessments. The training set consists of 688 speakers, while the development and evaluation sets include 119 speakers and 95 speakers, respectively. After silence stripping and data augmentation \cite{cuhk_elderly_zi_ye}, the size of the augmented training set expands to 58.9 hours, while the development and evaluation sets contain 2.5 hours and 0.6 hours of audio, respectively. The Cantonese \textbf{JCCOCC MoCA}\cite{jccocc_datasets} corpus 
comprises 256 cognitive impairment assessment interviews between elderly participants and clinical investigators. The training set contains 369 speakers, while the development and evaluation sets each include recordings from two different groups of 49 elderly speakers. \textbf{No elderly speakers in the training set overlap with those in the development or evaluation sets for both corpora.} 
\par
\noindent
\textbf{Experimental setup:}
We adopt Whisper-medium\footnote{https://huggingface.co/openai/whisper-medium} for its strong generalization ability, and apply LoRA\cite{lora} on the ``query'', ``key", ``value", and ``att.out" of the attention module, 
with the LoRA rank set to 8. \textbf{These fine-tuning results achieve state-of-the-art baseline performance on both datasets, outperforming other foundation models~\cite{ssl_shujie_taslp}.} 
For the router network, the global context module is constructed by stacking two standard multi-head attention layers, each followed by layer normalization, the downsampling network comprises three sequential blocks with progressively decreasing channel sizes (512, 256, and 128), followed by three linear layers that sequentially reduce the dimensionality to 1200, 1000, and 50. Dropout is applied after each fully-connected layer to prevent overfitting.

\begin{table}[htbp]
    \vspace{-0.15cm}
    \caption{Performance of batch-mode (Sys.B1-8) and online (Sys.O1-16) adaptation with different settings. The $\mathcal{L}_{ASR}$ is used by default in multi-task learning. Spkr loss denotes to $\mathcal{L}_{Spkr}$ and MSE loss represents the $\mathcal{L}_{MSE}$. ``Enc (Dec)" denotes Encoder (or Decoder) speaker prompts.}
    \centering
    \vspace{-0.2cm}
    \renewcommand{\arraystretch}{1.2} 
    \resizebox{\linewidth}{!}
    {
    \begin{tabular}{c|c|c|c|c|c|cc|cc|c}
    \hline\hline 
    \multirow{3}{*}{Sys.} & 
    \multicolumn{2}{c|}{Multi-task.} & 
    \multirow{3}{*}{\shortstack{Prompt\\ Pos.}} & 
    \multirow{3}{*}{\shortstack {Prompt\\Length}} & 
    \multirow{3}{*}{\shortstack {Prompt \\ Cluster}} & 
    \multicolumn{5}{c}{DementiaBank Pitt WER(\%)} \\
    \cline{2-3} \cline{7-11}
    &\multirow{2}{*}{\shortstack{MSE \\ loss}} & \multirow{2}{*}{\shortstack{Spkr\\loss 
    }} &&&&
    \multicolumn{2}{c|}{Dev.} & \multicolumn{2}{c|}{Eval.} & \multirow{2}{*}{All} \\
    \cline{7-10}
    &&&&&&Par. &Inv. &Par. &Inv. \\
    \hline\hline

B1&\multirow{4}{*}{\xmark}& \multirow{4}{*}{\xmark}&\multirow{4}{*}{Enc}&1&\multirow{4}{*}{\xmark}&27.83&12.62&19.48&11.88&19.77\\

B2&&&&2&&27.53&12.70&19.36&12.10&19.67\\

{\cellcolor{cyan!15}B3}&&&&4&&\textbf{27.53}&\textbf{12.44}&\textbf{19.53}&\textbf{12.43}&\textbf{19.60}\\

B4&&&&8&&27.96&11.94&19.82&12.32&19.61\\
\cline{1-11}

{\cellcolor{orange!20} B5}&\multirow{4}{*}{\xmark}& \multirow{4}{*}{\xmark}&\multirow{4}{*}{Dec}&1&\multirow{4}{*}{\xmark}&\textbf{28.12}&\textbf{12.16}&\textbf{20.01}&\textbf{12.76}&\textbf{19.81}\\

B6&&&&2&&28.95&12.47&20.30&12.54&20.31\\

B7&&&&4&&28.21&12.48&20.76&12.76&20.10\\

B8&&&&8&&28.47&12.65&20.76&12.76&20.28\\
\hline\hline

O1&{\xmark}&{\xmark}&\multirow{10}{*}{Enc} & \multirow{10}{*}{4}& \multirow{4}{*}{\shortstack {688\\(Unclustered)}}&28.69&12.77&19.95&12.10&20.25\\

O2&{\xmark}&{\cmark}&&&&28.78&12.84&20.03&12.54&20.35\\

O3&{\cmark}&{\xmark}&&&&28.76&12.57&19.90&11.65&20.18\\

O4&{\cmark}&{\cmark}&&&&28.69&12.68&19.74&11.99&20.18\\
\cline{2-3} \cline{6-11}

O5&{\cmark}&{\cmark}&&&25&28.92&12.40&19.67&12.10&20.15\\
\cline{2-3} \cline{6-11}

O6&{\xmark}&{\xmark}&&&\multirow{4}{*}{50}&28.69&12.80&19.82&12.65&20.26\\

O7&{\xmark}&{\cmark}&&&&29.17&12.45&19.86&12.76&20.32\\

O8&{\cmark}&{\xmark}&&&&28.73&12.67&19.61&10.99&20.14\\
{\cellcolor{yellow!50}O9}&{\cmark}&{\cmark}&&&&\textbf{28.49}&\textbf{12.32}&\textbf{19.48}&\textbf{10.88}&\textbf{19.88}\\
\cline{2-3} \cline{6-11}

O10&{\cmark}&{\cmark}&&&150&28.70&12.57&19.80&11.43&20.13\\
\hline\hline
\end{tabular}

}\label{tab:table_ablation}
\end{table}
\vspace{-0.3cm}

\begin{figure}[h]
    \centering
\includegraphics[width=0.4\textwidth]{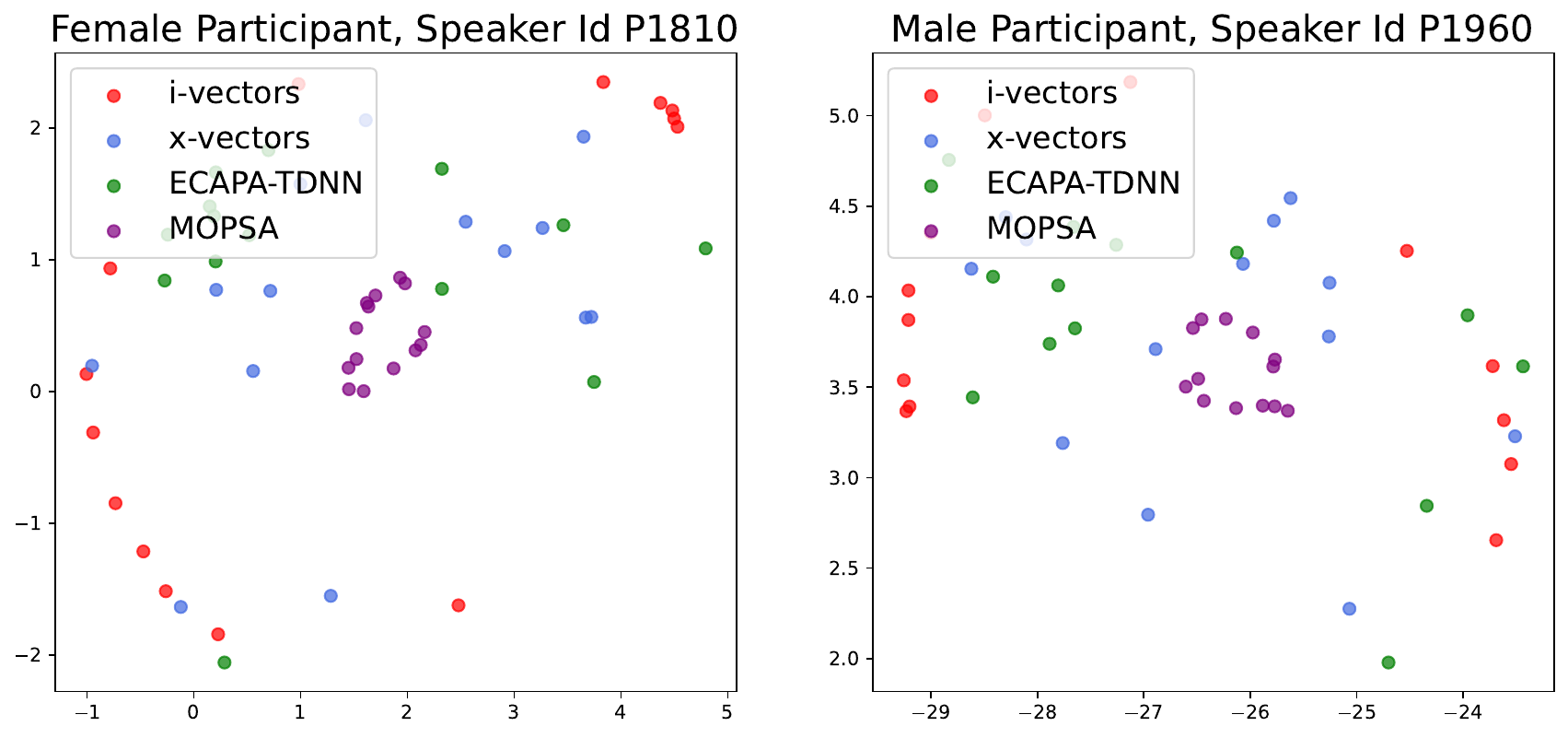}
    \caption{T-SNE visualization of the MOPSA, i/x-vectors and ECAPA-TDNN features (Sys.7-10, Tab. \ref{tab:table_online}) for 2 speakers in DementiaBank Eval set.  }
\label{fig:online_feature}
    \vspace{-0.4cm}
\end{figure}

\par
\begin{figure}[h]
    \centering
\includegraphics[width=0.25\textwidth]{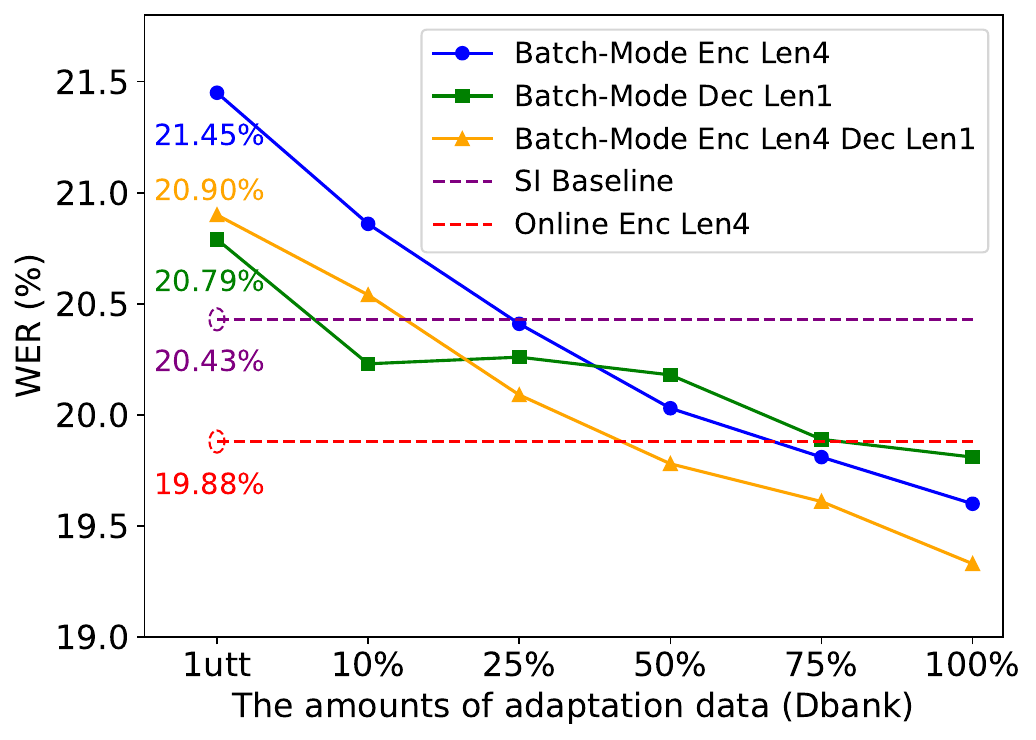}
    \caption{WER\% of adapted Whisper w.r.t. varying amounts of speaker data on the DementiaBank (Dev+Eval) data.}
    \label{fig:adapt_data}
    \vspace{-0.7cm}
\end{figure}
\begin{figure}[h]
    \centering
\includegraphics[width=0.38\textwidth]{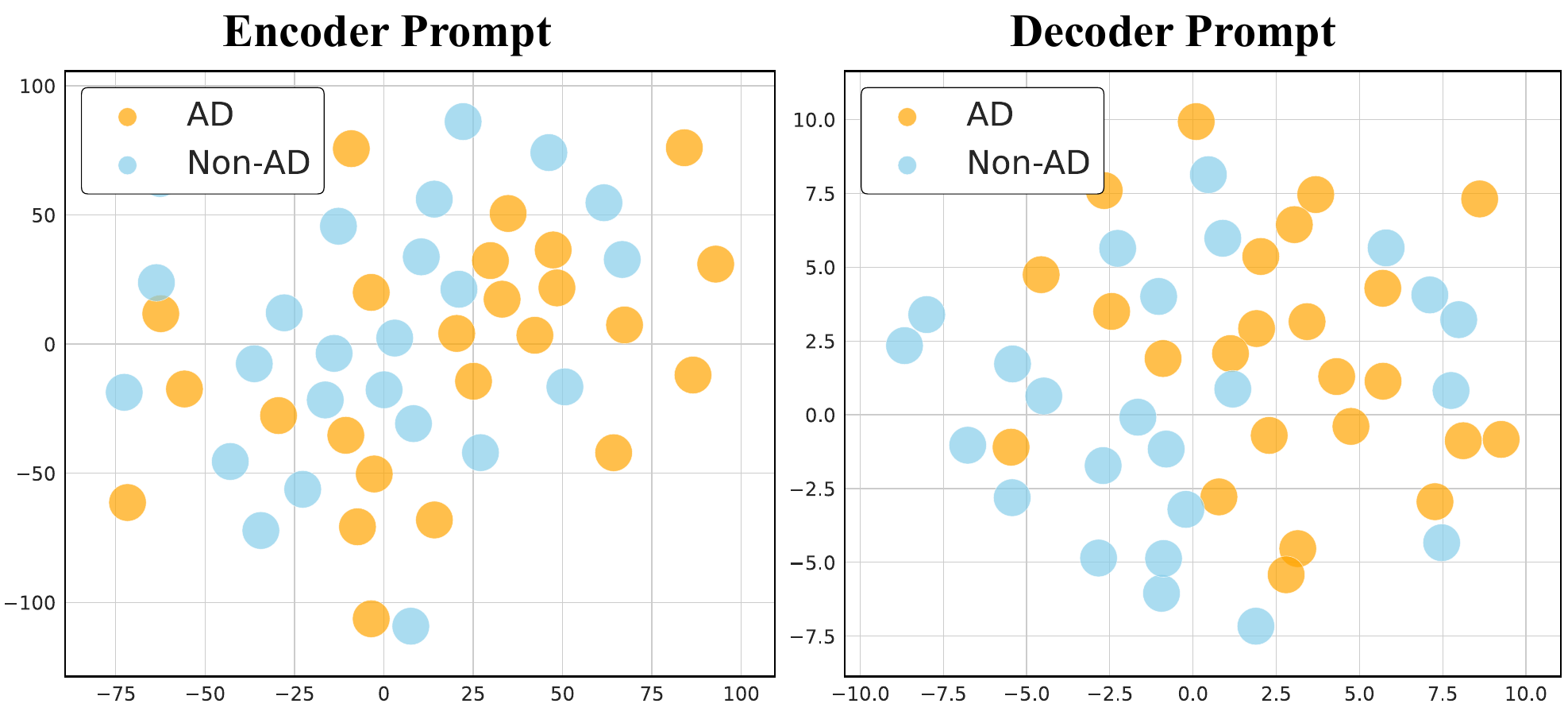}
     \caption{T-SNE visualization of batch-mode estimated encoder prompts (left, Sys 4, Tab. \ref{tab:table_online}) and decoder prompts (right, Sys 5, Tab. \ref{tab:table_online}) for 48 elderly speakers of DementiaBank Eval set. Speaker-level AD/non-AD labels in orange/blue.}
    \label{fig:decoder_prompt}
    \vspace{-0.4cm}
\end{figure}

\noindent
\textbf{Performance analysis:}  
As shown in Table \ref{tab:table_online}, 
several trends can be found: \textbf{1)} The superior batch-mode performance of jointly using encoder and decoder prompts, compared to using them individually (Sys.6 vs. 4,5), indicates that they capture complementary information. This combined approach significantly outperforms the baseline system with statistically significant WER reductions of up to  \textbf{1.10\% absolute (5.38\% relative)} on DementiaBank (Sys.6 vs. 1) and CER reductions of up to  \textbf{1.54\% absolute (5.66\% relative)} on JCCOCC MoCA (Sys.6 vs. 1). \textbf{2)} The online i/x-vectors ECAPA-TDNN underperforms against the SI baseline systems (Sys.7-9 vs. 1). While our MOPSA adaptation outperforms the baseline with statistically significant WER reductions of up to \textbf{0.55\% absolute (2.69\% relative)} on DementiaBank (Sys.10 vs. 1) and CER reductions of up to  \textbf{0.64\% absolute (2.35\% relative)} on JCCOCC MoCA (Sys.10 vs. 1). These gains are consistent with the T-SNE visualization in Fig. \ref{fig:online_feature} where speaker representations produced by online MOPSA adaptation are more consistent than those obtained from i/x-vectors and ECAPA-TDNN. \textbf{3)} Online MOPSA adaptation performance is comparable to that of batch-mode prompt-based adaptation on encoder-sid (Sys.10 vs. 4),  while offering \textbf{16.12} times speedup in inference RTF. Fig. \ref{fig:adapt_data} shows the performance of online MOPSA adaptation is largely invariant 
against speaker data quantity, in contrast to batch-mode adaptation (blue solid line vs. red dashed line). 
\textbf{4)} When applied separately to either the encoder or decoder, online MOPSA adaptation achieves performance comparable to that of batch-mode prompt-based adaptation (Sys.10 vs. 4, Sys.11 vs. 5, Sys.12 vs. 6).
\textbf{5)} As shown in Fig. \ref{fig:decoder_prompt}, the decoder prompts (Sys.5) intuitively correlate more strongly with Alzheimer’s disease diagnosis labels compared to the encoder prompts (Sys.4). 
This is consistent with the more important role of linguistic features for AD detection than acoustic features, as widely reported in prior researches \cite{wang2023exploiting,ssl_shujie_taslp,li2021comparative,wang22k_interspeech}. 

\par
\noindent
\textbf{Ablation study:} 
As shown in Table \ref{tab:table_ablation}, different configurations of batch-mode and online methods are investigated on DementiaBank Pitt. \textbf{For batch-mode adaptation},  
varying the lengths of prompts impacts model performance. The optimal results are obtained using the empirically set encoder and decoder prompt lengths of 4 and 1 (Sys.B3, B5). 
\textbf{For online adaptation}, the optimal settings use a multi-task criteria interpolation between the MSE and speaker recognition (``Spkr") losses (Sys.O6-O9), and 50 prompt clusters (Sys.O9 vs. 4,5,10). 
The above settings are used for all the main experiments in Table \ref{tab:table_online}.

\vspace{-0.3cm}
\section{Conclusion}
This paper proposes a novel Mixture of Prompt-Experts based Speaker Adaptation approach (MOPSA) for elderly speech recognition. Cluster-based prompt-experts and dynamic router networks allow zero-shot, real-time adaptation to unseen speakers. Experiments on DementiaBank Pitt and JCCOCC MoCA elderly speech datasets suggest that our method outperforms the SI Whisper model by WER and CER reductions of 0.86\% and 1.47\% absolute (4.21\% and 5.40\% relative) on the two tasks, respectively. Real-time factor (RTF) speed-up ratios of up to 16.12 times are obtained over offline batch-mode speaker prompts based adaptation.  

\section{Acknowledgements}
This research is supported by Hong Kong RGC GRF grant No. 14200220, 14200021, 14200324, Innovation Technology Fund grant No. ITS/218/21, China Disabled Persons Federation (CDPF2023KF00002), Basic Research Project of ISCAS (ISCAS-JCMS-202306), Youth Innovation Promotion Association CAS Grant (2023119), and Guangzhou CASTF project (2022MZK02).

\bibliographystyle{IEEEtran}
\bibliography{mybib}

\begin{thebibliography}{10}
\providecommand{\url}[1]{#1}
\csname url@samestyle\endcsname
\providecommand{\newblock}{\relax}
\providecommand{\bibinfo}[2]{#2}
\providecommand{\BIBentrySTDinterwordspacing}{\spaceskip=0pt\relax}
\providecommand{\BIBentryALTinterwordstretchfactor}{4}
\providecommand{\BIBentryALTinterwordspacing}{\spaceskip=\fontdimen2\font plus
\BIBentryALTinterwordstretchfactor\fontdimen3\font minus \fontdimen4\font\relax}
\providecommand{\BIBforeignlanguage}[2]{{%
\expandafter\ifx\csname l@#1\endcsname\relax
\typeout{** WARNING: IEEEtran.bst: No hyphenation pattern has been}%
\typeout{** loaded for the language `#1'. Using the pattern for}%
\typeout{** the default language instead.}%
\else
\language=\csname l@#1\endcsname
\fi
#2}}
\providecommand{\BIBdecl}{\relax}
\BIBdecl

\bibitem{becker1994natural_dbank}
J.~T. Becker, F.~Boiler, O.~L. Lopez, J.~Saxton, and K.~L. McGonigle, ``The natural history of alzheimer's disease: description of study cohort and accuracy of diagnosis,'' \emph{Archives of neurology}, vol.~51, no.~6, pp. 585--594, 1994.

\bibitem{whisper}
A.~Radford, J.~W. Kim, T.~Xu, G.~Brockman, C.~McLeavey, and I.~Sutskever, ``{Robust Speech Recognition via Large-Scale Weak Supervision},'' in \emph{ICML}, 2023.

\bibitem{baevski2020wav2vec}
A.~Baevski, Y.~Zhou, A.~Mohamed, and M.~Auli, ``wav2vec 2.0: A framework for self-supervised learning of speech representations,'' \emph{Advances in neural information processing systems}, 2020.

\bibitem{hsu2021hubert}
W.-N. Hsu, B.~Bolte, Y.-H.~H. Tsai, K.~Lakhotia, R.~Salakhutdinov, and A.~Mohamed, ``Hubert: Self-supervised speech representation learning by masked prediction of hidden units,'' \emph{IEEE/ACM T-ASLP}, 2021.

\bibitem{chen2022wavlm}
S.~Chen, C.~Wang, Z.~Chen, Y.~Wu, S.~Liu, Z.~Chen, J.~Li, N.~Kanda, T.~Yoshioka, X.~Xiao \emph{et~al.}, ``Wavlm: Large-scale self-supervised pre-training for full stack speech processing,'' \emph{IEEE Journal of Selected Topics in Signal Processing}, vol.~16, no.~6, pp. 1505--1518, 2022.

\bibitem{hu2022exploring}
S.~Hu, X.~Xie, Z.~Jin, M.~Geng, Y.~Wang, M.~Cui, J.~Deng, X.~Liu, and H.~Meng, ``{Exploring Self-supervised Pre-trained ASR Models For Dysarthric and Elderly Speech Recognition},'' in \emph{ICASSP}, 2023.

\bibitem{ssl_shujie_taslp}
S.~Hu, X.~Xie, M.~Geng, Z.~Jin, J.~Deng, G.~Li, Y.~Wang, M.~Cui, T.~Wang, and H.~Meng, ``Self-supervised asr models and features for dysarthric and elderly speech recognition,'' \emph{IEEE/ACM T-ASLP}, 2024.

\bibitem{geng2024homogeneous}
M.~Geng, X.~Xie, J.~Deng, Z.~Jin, G.~Li, T.~Wang, S.~Hu, Z.~Li, H.~Meng, and X.~Liu, ``Homogeneous speaker features for on-the-fly dysarthric and elderly speaker adaptation,'' \emph{arXiv preprint arXiv:2407.06310}, 2024.

\bibitem{geng2022speaker}
M.~Geng, X.~Xie, Z.~Ye, T.~Wang, G.~Li, S.~Hu, X.~Liu, and H.~Meng, ``{Speaker adaptation using spectro-temporal deep features for dysarthric and elderly speech recognition},'' \emph{IEEE/ACM T-ASLP}, 2022.

\bibitem{hu2024structured}
S.~Hu, X.~Xie, M.~Geng, J.~Deng, Z.~Jin, T.~Wang, M.~Cui, G.~Li, Z.~Li, H.~Meng \emph{et~al.}, ``Structured speaker-deficiency adaptation of foundation models for dysarthric and elderly speech recognition,'' \emph{arXiv preprint arXiv:2412.18832}, 2024.

\bibitem{cuhk_elderly_zi_ye}
Z.~Ye, S.~Hu, J.~Li, X.~Xie, M.~Geng, J.~Yu, J.~Xu, B.~Xue, S.~Liu, X.~Liu \emph{et~al.}, ``{Development of the CUHK Elderly Speech Recognition System for Neurocognitive Disorder Detection Using the Dementiabank Corpus},'' in \emph{ICASSP}, 2021.

\bibitem{gemello2007linear}
R.~Gemello, F.~Mana, S.~Scanzio, P.~Laface, and R.~De~Mori, ``{Linear hidden transformations for adaptation of hybrid ANN/HMM models},'' \emph{Speech Communication}, 2007.

\bibitem{li2010comparison}
B.~Li and K.~C. Sim, ``{Comparison of discriminative input and output transformations for speaker adaptation in the hybrid NN/HMM systems},'' in \emph{INTERSPEECH}, 2010.

\bibitem{tan2015cluster}
T.~Tan, Y.~Qian, and K.~Yu, ``{Cluster adaptive training for deep neural network based acoustic model},'' \emph{IEEE/ACM T-ASLP}, 2015.

\bibitem{zhang2016dnn}
C.~Zhang and P.~C. Woodland, ``{DNN speaker adaptation using parameterised sigmoid and ReLU hidden activation functions},'' in \emph{ICASSP}, 2016.

\bibitem{zhao2016low}
Y.~Zhao, J.~Li, and Y.~Gong, ``{Low-rank plus diagonal adaptation for deep neural networks},'' in \emph{ICASSP}, 2016.

\bibitem{samarakoon2016subspace}
L.~Samarakoon and K.~C. Sim, ``Subspace lhuc for fast adaptation of deep neural network acoustic models.'' in \emph{INTERSPEECH}, 2016.

\bibitem{li2018speaker}
K.~Li, J.~Li, Y.~Zhao, K.~Kumar, and Y.~Gong, ``Speaker adaptation for end-to-end ctc models,'' in \emph{SLT}, 2018.

\bibitem{xie2019fast}
X.~Xie, X.~Liu, T.~Lee, and L.~Wang, ``Fast dnn acoustic model speaker adaptation by learning hidden unit contribution features.'' in \emph{INTERSPEECH}, 2019.

\bibitem{fan2019speaker}
Z.~Fan, J.~Li, S.~Zhou, and B.~Xu, ``Speaker-aware speech-transformer,'' in \emph{ASRU}, 2019.

\bibitem{huang2021rapid}
Y.~Huang, G.~Ye, J.~Li, and Y.~Gong, ``Rapid speaker adaptation for conformer transducer: Attention and bias are all you need.'' in \emph{INTERSPEECH}, 2021.

\bibitem{sim2021robust}
K.~C. Sim, A.~Chandorkar, F.~Gao, M.~Chua, T.~Munkhdalai, and F.~Beaufays, ``Robust continuous on-device personalization for automatic speech recognition.'' in \emph{INTERSPEECH}, 2021.

\bibitem{deng2023confidence}
J.~Deng, X.~Xie, T.~Wang, M.~Cui, B.~Xue, Z.~Jin, G.~Li, S.~Hu, and X.~Liu, ``Confidence score based speaker adaptation of conformer speech recognition systems,'' \emph{IEEE/ACM TASLP}, 2023.

\bibitem{jiajun_high_performance}
J.~Deng, X.~Xie, G.~Li, M.~Cui, M.~Geng, Z.~Jin, T.~Wang, S.~Hu, Z.~Li, and X.~Liu, ``{Towards High-Performance and Low-Latency Feature-Based Speaker Adaptation of Conformer Speech Recognition Systems},'' in \emph{ICASSP}, 2024.

\bibitem{jacobs1991adaptive_moe}
R.~A. Jacobs, M.~I. Jordan, S.~J. Nowlan, and G.~E. Hinton, ``{Adaptive mixtures of local experts},'' \emph{Neural computation}, 1991.

\bibitem{jordan1994hierarchical_moe}
M.~I. Jordan and R.~A. Jacobs, ``{Hierarchical mixtures of experts and the EM algorithm},'' \emph{Neural computation}, 1994.

\bibitem{lora}
E.~J. Hu, Y.~Shen, P.~Wallis, Z.~Allen-Zhu, Y.~Li, S.~Wang, L.~Wang, and W.~Chen, ``{LoRA: Low-Rank Adaptation of Large Language Models},'' in \emph{ICLR}, 2022.

\bibitem{adapter_tuning}
N.~Houlsby, A.~Giurgiu, S.~Jastrzebski, B.~Morrone, Q.~De~Laroussilhe, A.~Gesmundo, M.~Attariyan, and S.~Gelly, ``{Parameter-efficient transfer learning for NLP},'' in \emph{ICML}, 2019.

\bibitem{zhao24_saml_interspeech}
Q.~Zhao, G.~Sun, C.~Zhang, M.~Xu, and T.~F. Zheng, ``Saml: Speaker adaptive mixture of lora experts for end-to-end asr,'' in \emph{INTERSPEECH}, 2024.

\bibitem{liu2021gpt}
X.~Liu, Y.~Zheng, Z.~Du, M.~Ding, Y.~Qian, Z.~Yang, and J.~Tang, ``{GPT Understands, Too},'' \emph{arXiv:2103.10385}, 2021.

\bibitem{jiang24b_interspeech}
Y.~Jiang, T.~Wang, X.~Xie, J.~Liu, W.~Sun, N.~Yan, H.~Chen, L.~Wang, X.~Liu, and F.~Tian, ``{Perceiver-Prompt: Flexible Speaker Adaptation in Whisper for Chinese Disordered Speech Recognition},'' in \emph{INTERSPEECH}, 2024.

\bibitem{anastasakos1996compact_sat}
T.~Anastasakos, J.~McDonough, R.~Schwartz, and J.~Makhoul, ``{A compact model for speaker-adaptive training},'' in \emph{ICSLP}, 1996.

\bibitem{jccocc_datasets}
S.~S. Xu, M.-W. Mak, K.~H. Wong, H.~Meng, and T.~C. Kwok, ``Speaker turn aware similarity scoring for diarization of speech-based cognitive assessments,'' in \emph{APSIPA ASC}, 2021.

\bibitem{gillick1989some}
L.~Gillick and S.~J. Cox, ``{Some statistical issues in the comparison of speech recognition algorithms},'' in \emph{ICASSP}, 1989.

\bibitem{wang2024advancing}
S.~Wang, Z.~Chen, B.~Han, H.~Wang, C.~Liang, B.~Zhang, X.~Xiang, W.~Ding, J.~Rohdin, A.~Silnova \emph{et~al.}, ``{Advancing speaker embedding learning: Wespeaker toolkit for research and production},'' \emph{Speech Communication}, 2024.

\bibitem{attention}
A.~Waswani, N.~Shazeer, N.~Parmar, J.~Uszkoreit, L.~Jones, A.~Gomez, L.~Kaiser, and I.~Polosukhin, ``{Attention is All you Need},'' in \emph{{NeurIPS}}, 2017.

\bibitem{wang2023exploiting}
Y.~Wang, J.~Deng, T.~Wang, B.~Zheng, S.~Hu, X.~Liu, and H.~Meng, ``Exploiting prompt learning with pre-trained language models for alzheimer’s disease detection,'' in \emph{ICASSP}, 2023.

\bibitem{li2021comparative}
J.~Li, J.~Yu, Z.~Ye, S.~Wong, M.~Mak, B.~Mak, X.~Liu, and H.~Meng, ``A comparative study of acoustic and linguistic features classification for alzheimer's disease detection,'' in \emph{ICASSP}, 2021.

\bibitem{wang22k_interspeech}
T.~Wang, J.~Deng, M.~Geng, Z.~Ye, S.~Hu, Y.~Wang, M.~Cui, Z.~Jin, X.~Liu, and H.~Meng, ``Conformer based elderly speech recognition system for alzheimer’s disease detection,'' in \emph{INTERSPEECH}, 2022.

\end{thebibliography}

\end{document}